\documentclass[12pt]{article}
\newlength{\mytopmargin}
\newlength{\myleftmargin}
\usepackage{amsmath,amsthm,amssymb,amsbsy,epsfig,graphicx,color,multicol,subfigure}
\usepackage{array,calc}
\usepackage[enableskew]{youngtab}
\usepackage{rotating}
\usepackage{young,epic}
\usepackage{a4wide,bm}
\usepackage{url}
\usepackage{hyperref}

\newtheorem{theorem}{Theorem}

\usepackage{amsmath,amsfonts,amssymb}

\usepackage{graphicx}

\begin{document}
%

\title{Local Central Limit Theorem for Determinantal Point Processes}
\author{Peter J. Forrester${}^\dagger$ and Joel L. Lebowitz${}^*$}
\date{}
\maketitle
\noindent
\thanks{\small ${}^\dagger$Department of Mathematics and Statistics, 
The University of Melbourne,
Victoria 3010, Australia email:  p.forrester@ms.unimelb.edu.au \\
${}^*$Departments of Mathematics and Physics, Rutgers University, NJ 08854-8019, USA
email: lebowitz@math.rutgers.edu; Institute for Advanced Study, Princeton, NJ 08540, USA
}

\begin{abstract}
\noindent We prove a local central limit theorem (LCLT) for the number of points $N(J)$
in a region
$J$ in $\mathbb R^d$ specified by a determinantal point process
with an Hermitian kernel. The only assumption is
that the variance of $N(J)$ tends to infinity as $|J| \to \infty$. 
This extends a previous result giving a weaker central limit theorem (CLT)
for these systems. 
Our result relies on the fact that 
the Lee-Yang zeros of the generating function for  $\{E(k;J)\}$ --- the probabilities
of there being exactly $k$ points in $J$ --- all lie on the negative real $z$-axis. 
 In particular, the result applies to
the scaled bulk eigenvalue distribution for the Gaussian Unitary Ensemble (GUE) and that of the
Ginibre ensemble. For the GUE we can also treat the properly scaled edge eigenvalue distribution.
Using identities between gap probabilities,
the LCLT can be extended to bulk eigenvalues of the Gaussian Symplectic Ensemble (GSE).
A LCLT is also established for the probability density function of the $k$-th largest eigenvalue
at the soft edge,
and of the spacing between $k$-th neigbors in the bulk.

\end{abstract}

\section{Introduction}

Determinantal point processes are prominent structures in the theory of random matrices
as well as many other contexts \cite{So00}.
These are processes for which the
$k$-point correlation function can be written as a $k \times k$ determinant,
\begin{equation}\label{rK}
\rho_{(k)}(x_1,\dots,x_k) = \det [ K(x_j, x_l) ]_{j,l=1,\dots,k},
\end{equation}
where $K(x,y)$ --- referred to as the correlation kernel --- is independent of $K$.
A necessary and sufficient condition for (\ref{rK}) to represent a point process in $J$, when $K$
(viewed as the kernel for an integral operator supported on $J$) is Hermitian, is that all
its eigenvalues be discrete and lie between zero and one  (see e.g.\cite{HKPV08}).
Such $K$'s are the only one we shall consider here.

One of the best known examples of a determinantal point process is given by the eigenvalues of the random matrices specified by the GUE: a Gaussian probability measure on the space of 
complex $N \times N$ Hermitian matrices which is unitary invariant and thus unchanged by
conjugation by unitary matrices (see e.g.~\cite{Fo10,PS11}).
By scaling the eigenvalues so that the mean density is unity and taking $N \to \infty$, one obtains a translation invariant determinantal point process specified by the
so-called sine
kernel $K(x,y) = \sin \pi (x - y)/
\pi (x - y)$. The GUE also admits a soft edge scaling in the neighborhood of the largest eigenvalue, 
which now involves changing the origin so that it is centered near the largest eigenvalues, then scaling so the expected spacing between eigenvalues in this neighbourhood is of order unity in
the limit $N \to \infty$.
This is called a soft edge scaling
since $x=0$ is a soft wall --- eigenvalues do occur in the region $x>0$ but  their
density falls off super-exponentially. The resulting point process defined by the eigenvalues is determinantal with the explicit form of the correlation kernel given by
the Airy kernel
$K(x,y) = ({\rm Ai}(x) {\rm Ai}'(y) - {\rm Ai}(y) {\rm Ai}'(x) )/(x-y)$.

The eigenvalues of the Ginibre ensemble of non-Hermitian matrices with standard complex entries
give  an example of a 
determinantal point process with a complex Hermitian kernel: in the limit $N \to \infty$
this  is given by $K(w,z) = {1 \over \pi} e^{-(|w|^2+|z|^2)/2}
e^{w \bar{z}}$ (see e.g.~\cite{Me91}), where $z$ and  $w$ are complex.

It is the purpose of the present Letter to give a local central limit theorem (LCLT)
for the probabilities $E(k;J)$ that there are exactly $k$ points in  $J$, where $k$ is close to the expected number of points in $J$, 
in the limit $|J| \to
\infty$, for the class of determinantal
point processes introduced in the first paragraph. We begin in Section \ref{S2}
by recalling the  central limit theorem (CLT) of Costin and Lebowitz \cite{CL95}  for number fluctuations in determinantal point processes, and then 
giving an alternative  derivation which uses only  the location of the eigenvalues  of the
underlying integral operator, or equivalently the zeros of the generating function for
$\{E(k;J)\}$.
We will then prove the LCLT by using a theorem of Newton to establish
log-concavity of $\{E(k;J\})$, which is a known sufficient condition for a CLT to imply a
LCLT. In Section \ref{S3}
 the LCLT theorem
is  applied to specify the distribution of $E(k;J)$
for the scaled limits, both bulk and soft edge, of the GUE,
and for the Ginibre ensemble of non-Hermitian complex
random matrices (see e.g.~\cite{Me91} for precise definitions).
We extend the results for the GUE to the
GSE and in part also to the GOE (Gaussian  symplectic and
orthogonal ensemble) by making use
of inter-relation formulas from \cite{Me92}.
We furthermore obtain a LCLT for the distribution of the $k$-th largest
eigenvalue at the soft edge, and the distribution of the spacing between $k$-th
neighbors in the bulk.

\section{A local limit theorem}\label{S2}
\setcounter{equation}{0}
Our setting is a determinantal point process in $\mathbb R^d$. We denote by $N(J)$ the
random variable for the number of points in $J \subset \mathbb R^d$. We set
$\mu_J = {\rm mean} \, (N(J))$ and $\sigma_J^2 = {\rm Var} \, (N(J))$, and we
denote by $E(k;J)$ the probability that there are exactly $k$ points in $J$. We remark
that in terms of the correlation functions
\begin{equation}\label{JJ}
\mu_J = \int_J \rho_{(1)}(x) \, dx, \qquad
\sigma_J^2 = \int_J dx_1  \int_J dx_2 \, ( \rho_{(2)}^T(x_1,x_2) +
\rho_{(1)}(x_1) \delta (x_1 - x_2)).
\end{equation}

Costin and Lebowitz \cite{CL95} studied $N(J)$ for
the particular determinantal point process corresponding to the 
eigenvalues of the GUE in the limit $N \to \infty$, scaled 
 so that $\mu_J = |J|$ (bulk scaling limit) and thus specified by the sine kernel.
 They proved the CLT
\begin{equation}\label{rK1a}
\lim_{|J| \to \infty} {(N(J) - \mu_J) \over \sigma_J} \mathop{=}\limits^{\rm d} \eta,
\end{equation}
where $\eta$ is a standard Gaussian random variable. This was done by showing that
 as a consequence of the property that $\sigma_J \to \infty$ as $|J| \to \infty$,
all cumulants of the characteristic function beyond the second vanish for
$|J| \to \infty$.
In fact the proof makes no explicit use of the particular determinantal point process under
consideration, requiring only that the corresponding kernel be locally trace class
and self-adjoint,
and that
the variance tends to infinity, and so (\ref{rK1a}) 
is a universal property of determinantal
point processes in this setting
 (see also \cite{So00c}).

Important for our study of the LCLT is a derivation of the CLT based on the
generating function (grand partition function)
\begin{equation}\label{rK5}
\Xi(z;J) = \sum_{k=0}^\infty z^k E(k;J).
\end{equation}
Generally \cite[Eq.~(9.4)]{Fo10}
\begin{equation}\label{rK6}
\Xi(1 - \xi ;J)  = 1 + \sum_{k=1}^\infty {(-1)^k  \over k!} \xi^k
\int_J d x_1 \cdots  \int_J d x_d \, \rho_{(k)}(x_1,\dots,x_k) .
\end{equation}
In the case of a determinantal point process, and thus $\rho_{(k)}$ given by (\ref{rK}), it is well
known (see e.g.~\cite{WW65}) that the sum on the RHS is the expanded form of the
determinant of the Fredholm integral operator with kernel $K$ supported on $J$, and thus can be written
\begin{equation}\label{rK7}
\Xi(1 - \xi ;J)  =  \prod_{l=0}^\infty (1 - \xi \lambda_l(J)),
\end{equation}
where the $\lambda_l(J)$ are eigenvalues of the integral operator $K$ supported on $J$.
Equivalently, with $C$ independent of $z$,
\begin{equation}\label{rK8}
\Xi(z ;J)  =  C \prod_{l=0}^\infty (1 + z \mu_l(J)), \qquad \mu_l(J) = {\lambda_l(J) \over 1 - \lambda_l(J)}
\end{equation}
Restricting attention to kernels such that the integral operator is
locally trace class and self-adjoint implies that
\begin{equation}\label{rK9}
1 > \lambda_0(J) \ge \lambda_1(J) \ge \cdots \ge  \lambda_n(J) \ge \cdots \ge 0,
\end{equation}
or equivalently
$
0 < \mu_0(J) < \mu_1(J) < \mu_2(J) < \cdots.
$
As remarked in \cite[proof of Theorem 4.6.1]{HKPV08}, and in fact observed much earlier
\cite{Ha67}, $N(J)$  can be viewed as the  sum of independent
but not identically distributed Bernoulli random variables $x_l \in \{0,1\}$
with Pr${}(x_l = 1) = \lambda_l(J)$. 
It follows then from the works of Harper \cite{Ha67} and Canfield \cite{Ca80} (see
 \cite[Section XVI.5, Theorem 2]{Fe66} )
that we have a CLT whenever
$\sigma_J^2  = \sum_{j=0}^\infty \lambda_j(J) (1 - \lambda_j(J))  \to \infty$ for
$J = J_s$ with $s \to \infty$. In view of the relationship between $N(J)$ and $\{E(k;J)\}$  this CLT can be written
\begin{equation}\label{rK4}
\lim_{s \to \infty} \, \mathop{\sup}\limits_{x \in (-\infty, \infty)}
\Big | \sum_{k \le \sigma_{J_s} x + \mu_{J_s}} E(k;J_s) -
{1 \over \sqrt{2 \pi}}  \int_{-\infty}^x e^{-t^2/2} \, dt \Big | = 0.
\end{equation}
Here we have introduced  a parameter $s$ in specifying the region $J$ so as to be able to consider the natural
interval $J = (-s,\infty)$ in the 
soft edge scaling case, which has $|J| = \infty$. For bulk scaling we can take $J_s = (0,s)$.

Stronger than (\ref{rK4})  would be a limiting form for
$E(k;J)$. This is provided by the following LCLT.

\begin{theorem}\label{T1}
Consider a determinantal point process labelled by a parameter $s$, and consider
a region $J = J_s$.  Suppose that 
the eigenvalues of the integral operator corresponding the correlation kernel $K(x,y)$
supported on $J_s$ are discrete and between 0 and 1 as in (\ref{rK9}), and that
$\sigma_{J_s}\to \infty$ as $s \to \infty$.
We then have that the
$E(k;J)$ satisfy the LCLT 
\begin{equation}\label{rK1}
\lim_{s \to \infty} \, \mathop{\sup}\limits_{x \in (-\infty, \infty)}
\Big | \sigma_{J_s} E([\sigma_{J_s} x + \mu_J]) - {1 \over \sqrt{2 \pi}} e^{- x^2/2} \Big | = 0.
\end{equation}
\end{theorem}

A condition for the passage from a central to a local limit theorem has been given
by Bender \cite{Be73}. All that is required  is that all the
zeros of $\Xi(z;J)$ are on the negative real axis. 
The proof that  this is sufficient goes via the fact that the restriction of zeros to the negative $z$-axis implies,
by Newton's theorem (see e.g.~\cite{Ni00}), that the $E(k;J)$ are log concave, i.e.
$
\log E(k+1;J) - 2 \log E(k;J) + \log E(k-1;J) \le 0$.
It is this latter property which is shown in
 \cite{Be73} to be a sufficient condition for the passage from a central to a local limit theorem.
 As noted above, the assumption  (\ref{rK9}) implies that the $\{\mu_k(J)\}$ are all positive real and thus
that the zeros of  $\Xi(z;J)$  are all on the negative real axis, thus establishing the validity
of Theorem \ref{T1}.

\section{Random matrix applications}\label{S3}
\setcounter{equation}{0}
\subsection*{Bulk GUE}
Perhaps the best known example of a determinantal point process is the bulk scaled GUE.
 In this limit the
correlations are given by (\ref{rK}) with the sine kernel.
For this kernel it is a readily derived consequence of the second formula in
(\ref{JJ}) (see e.g.~\cite[\S 14.5.1]{Fo10}) that for
large $|J|$, $\sigma_J^2 \sim (1/\pi^2) \log |J| + C/\pi^2 +
(1 + \log 2 \pi)/\pi^2$, where $C$ denotes Euler's constant.  In particular this diverges for
$|J| \to \infty$, or equivalently for $s \to \infty$ with $J=(0,s)$,
 so Theorem \ref{T1} applies.
The sine kernel is one of a whole class of kernels for which high precision
computation of the $E(k;J)$ is available using Bornemann's Matlab software \cite{Bo09}
based on the
Fredholm determinant formula (\ref{rK7}). 
Thus for a given finite $|J|$ we can compute the deviation of
the exact value from the limiting Gaussian form. This  is done in Table \ref{T1b}.
Note that this deviation is small, differing only in the third nonzero digit  for values of $k$ within
3 standard deviations of the mean.

\begin{table}
\begin{center}
\begin{tabular}{c||c|c|c|c|c|c|c}
$k$& 7 & 8 & 9 & 10 & 11 & 12 & 13  \\\hline
exact & $1.49 \times 10^{-4} $& 0.0161 & 0.2238 & 0.5202 & 0.2234 & 0.0163 & $1.61 \times
10^{-4}$ \\ \hline
Gaussian & $2.2 \times 10^{-4}$ & 0.0166 & 0.221 & 0.524 & 0.221 & 0.0166 & $2.2 \times
10^{-4}$
\end{tabular}
\caption{\label{T1b} Tabulation of $E(k;J)$ for the bulk GUE, with $|J| = 10$, and the corresponding
Gaussian form. In the latter $\mu_J = 10$ and $\sigma_J = 0.761$.}
\end{center}
\end{table}

\subsection*{Soft edge GUE}
The GUE also admits a soft edge scaling $\lambda \mapsto \sqrt{2N} + \lambda/(\sqrt{2} N^{1/6})$,
$N \to \infty$. This has the effect of moving the origin to the neighbourhood of the largest eigenvalue, and making the mean spacing in this neighbourhood of order unity.
The corresponding correlations are then given by (\ref{rK}) with $K(x,y)$ equal to the Airy kernel as specified in the second paragraph. The corresponding probability of there being $k$ eigenvalues in $J$ is denoted
$E^{\rm soft}(k;J)$, with the natural choice of $J$ being $ (-s, \infty)$. The Airy kernel is real symmetric and asymptotically $\sigma_J^2 = (1/2 \pi^2)\log s^{3/2}$ \cite[eq.~(2.30)]{Fo12a}),
so according to Theorem \ref{T1} $E^{\rm soft}(k;J)$ must obey the LCLT (\ref{rK1}).
Since the density of the eigenvalues at the soft edge has the asymptotic form
$|\lambda|^{1/2}/\pi$ for $\lambda \to - \infty$  \cite[eq.~(7.69)]{Fo10} one has
$\mu_J \sim  2 s^{3/2}/(3 \pi) + O(1)$, which together with the asymptotic form of $\sigma_J^2$
is data to be substituted into (\ref{rK1}). If we
take $s =  (15 \pi)^{2/3}$ so that $\mu_j \approx 10$,
Bornemann's package gives $E^{\rm soft}(10;(- (15 \pi)^{2/3},\infty)) = 0.6405$,
while the Gaussian form with $\mu_J = 9.99$, $\sigma_J^2 = 0.377$ as computed from
(\ref{JJ})  for this choice of $J$
gives $E^{\rm soft}(10;(- (15 \pi)^{2/3},\infty)) \approx 0.649$.
This shows that the limiting Gaussian
form is quite accurate even for small $\sigma_{J_s}$.

\subsection*{Conditioning with fixed eigenvalues}

For the scaled GUE and some other point processes in one-dimension one can define
the probability densities $\{p^{\rm soft}(k;(-s,\infty))\}$
for there being a particle at $-s$ and exactly $k$ particles in $(-s,\infty)$
(for this to make sense the soft wall at $x=0$ must be such that the expected number of particles
in $x>0$ is finite). Equivalently this is the probability density function for the distribution of the
$(k+1)$th largest eigenvalue.  One can also define
$\{p^{\rm bulk}(k;s)\}$ for there being exactly $k$ particles between two particles at 
separation $s$  in the bulk . In the case of the scaled GUE these probability densities also fit into the setting of Theorem \ref{T1} and thus satisfy a LCLT, as we will now demonstrate.
Equivalently $\{p^{\rm bulk}(k;s)\}$ is the probability density function for the $k$th neighbour spacing in the bulk.
 
 Consider first $p^{\rm soft}(k;(-s,\infty))$ in the case of the soft edge scaled GUE. We can view this as a gap probability in the soft edge GUE, conditioned to have an eigenvalue at $-s$. In such a setting,
 it is known \cite{IKO08,FW12} that  the correlation kernel, to be denoted
$K_s^{\rm soft}$, can be written in terms of the usual soft edge scaled GUE kernel according to
\begin{equation}\label{D}
K_s^{\rm soft}(x,y) = K^{\rm soft}(x,y) - { K^{\rm soft}(x,-s)  K^{\rm soft}(-s,y) \over  K^{\rm soft}(-s,-s)}.
\end{equation}
With $\Xi^{\rm soft}_s(z;J) := \sum_{k=0}^\infty z^k E_s(k;J)$, the conditions for the validity of
the LCLT are met provided $\sigma_{J_s} \to \infty$ as $s \to  \infty$. In the special case $J_s=(-s,\infty)$, and with $\rho^{\rm soft}(x)$ denoted the soft edge eigenvalue density, it follows from the
definitions that $E_s(k;J_s) = p^{\rm soft}(k;(-s,\infty))/\rho^{\rm soft}(-s)$, and so with $\mu_{J_s}$
and $\sigma_{J_s}^2$ taking the asymptotic values specified in the second paragraph of this section
we obtain
\begin{equation}\label{BB}
\lim_{s \to  \infty} \sigma_{J_s}  p^{\rm soft}([\sigma_{J_s} x + \mu_{J_s}];(-s,\infty))/\rho_{(1)}^{\rm soft}(-s) =
{1 \over \sqrt{2 \pi}} e^{- x^2/2}.
\end{equation}
If we set $\ell =[\sigma_{J_s} x + \mu_J]$, then for large $-s$, $-s \sim \mu_\ell - \sigma_\ell x$ with
$\mu_\ell = (3 \pi \ell /2)^{2/3}$ and $\sigma_\ell = \sigma_{J_s}/\rho_{(1)}^{\rm soft}(-s)$,  telling us
that (\ref{BB}) can be rewritten as the statement 
\begin{equation}\label{BB1}
\lim_{\ell \to \infty} \sigma_\ell p^{\rm soft}(\ell; ( -\mu_\ell + \sigma_\ell x,\infty)) =
{1 \over \sqrt{2 \pi}} e^{- x^2/2},
\end{equation}
which is a LCLT with respect to the continuous variable in the probability density. This latter
limit formula  is consistent with an analogous result for the fluctuations of the distribution of the 
$k$-th largest eigenvalue in the
finite $N$ GUE \cite{Gu05}, and extended to the GOE and GSE in \cite{OR10} (see also
\cite{BEY13}).

The reasoning required to establish a LCLT for $\{p^{\rm bulk}(k;s)\}$ is analogous. 
Denote by $K^{\rm bulk}_0(x,y)$ the correlation kernel for the bulk state at $\beta = 2$,
conditioned to have an eigenvalue at the origin. This is given by a certain Bessel kernel
(see \cite[eq.~(7.48)]{Fo10} with $a=1$). With $\rho_{(n)}^{0,s}$ ($\rho_{(n)}^0$) denoting
$n$-point correlation functions for the the bulk state conditioned to have eigenvalues at 0 and $s$ (at $0$) we have
$$
\rho_{(n+1)}^{0}(x_1,\dots,x_n,s) / \rho_{(1)}^{0}(s) =
\rho_{(n)}^{0,s}(x_1,\dots,x_n),
$$
so proceeding as in the derivation of (\ref{D}) gives
$$
K^{\rm bulk}_{0,s}(x,y) =   K^{\rm bulk}_{0}(x,y)  -
{  K^{\rm bulk}_{0}(x,s)   K^{\rm bulk}_{0}(s,y)  \over   K^{\rm bulk}_{0}(s,s)  }.
$$
The variance for large $s$ is determined by $ K^{\rm bulk}_{0}(x,y)$, and its variance
for large $s$ is determined by $K^{\rm bulk}(x,y)$, telling us in particular that the variance
diverges logarithmically in this limit. It follows that
$E_{0,s}(k;J)$ satisfies a local limit theorem. But $E_{0,s}(k;J) = p^{\rm bulk}(k;s) /
\rho_{(2)}^{\rm bulk}(0,s)$ thus giving a LCTL for the latter ratio, and furthermore
$\rho_{(2)}^{\rm bulk}(0,s) \to 1$ as $s \to \infty$ so the LCLT applies to $p^{\rm bulk}(k;s)$
itself. Moreover, the analogous change of variables in going from (\ref{BB}) to
(\ref{BB1}) implies that this can equivalently be regarded as a LCLT for the continuous
variable in the probability density with $\sigma_k = \sigma_J$. Heuristics and graphical evidence for such a limit theorem dates back to the early literature on random matrix theory
\cite[Appendix N, Fig.~9]{BFPW81}.

\subsection*{Ginibre ensemble}

The eigenvalues of random matrices also provide examples of a determinantal point process
in the plane (see e.g.~\cite[Ch.~15]{Fo10}). In an appropriate scaled $N \to \infty$ limit
these all give rise to the complex Hermitian kernel for the
the Ginibre ensemble of non-Hermitian standard complex Gaussian
matrices, mentioned in the second paragraph of the Introduction.
From \cite{MY80} we know that 
whenever $J$ can be generated 
from a fixed region $J_0$ by dilation, $\sigma_J^2 \sim -(|\partial J|/\pi)
\int_{\mathbb R^2} d\vec{r} \, |\vec{r}| \rho_{(2)}^T(\vec{r},\vec{0}) =
|\partial J|/(2 \pi^{3/2})$, where $|\partial J|$ denotes
the length of the perimeter of $J$, and the final equality follows from the explicit formula
$ \rho_{(2)}^T(\vec{r},\vec{0})  = - e^{- | \vec{r}|^2}$ as implied by the correlation kernel.
In particular $\sigma_J$ diverges as $|J| \to \infty$ so the LCLT  (\ref{rK1}) must hold.
In addition to the asymptotic value of $\sigma_J^2$ as noted, we furthermore have
$\mu_J = |J|/\pi^2$ as data in the LCLT.
The fast decay
of the truncated correlations allows the number density CLT  to be
studied using different methods \cite{MY80}, and furthermore extended to the case of
multiple neighbouring regions \cite{Le83}. However, we don't know of any alternative way
to derive the LCLT. In the case that $J$ is a disk, the eigenvalues in
(\ref{rK7}) are known explicitly (see e.g.~\cite[Prop.~15.5.3]{Fo10}), however this does not persist for other shaped regions,
nor is there an efficient numerical scheme to compute the corresponding Fredholm determinant.
We remark that in the case of lattice gases there are techniques which allow
a LCLT to be established without consideration of the Lee-Yang zeros
\cite{DT77}. 

\subsection*{GOE and GSE}
We now turn our attention to the bulk scaled GOE and GSE.
The statistical states formed by the eigenvalues
are examples of Pfaffian point processes (see e.g.~\cite[Ch.~6]{Fo10}). A formula analogous
to (\ref{rK7}) applies for the square of the generating function, 
$$
\Big ( \Xi(1 - \xi;J) \Big )^2 =  \det ( I - \xi J^{-1}A), \qquad J = \begin{bmatrix} 0 & 1 \\ -1 & 0
\end{bmatrix}
$$
where $A$ is a real $2 \times 2$ antisymmetric integral operator. As the $2 \times 2$ matrix integral operator
$J^{-1}A$ is not  self adjoint, we have no immediate information as to the location of the
zeros of the generating function. We note however that whenever the zeros of $\Xi(z;J)$
come in complex conjugate paris whose real parts are non-positive, and $\sigma_J \to
\infty$ then the process satisfies a CLT \cite{LPRS14}.

Independent of the location of the zeros of the generating function for the GOE and GSE
it was shown
in  \cite{CL95}  that the CLT (\ref{rK1a}) for the bulk scaled GUE implies
a CLT for the bulk scaled GOE and GSE 
(the GOE, GUE and GSE correspond to $\beta = 1,2,4$ in the Dyson-Mehta scheme and will
be so referred to below).
This was done by
using the facts that superimposing two GOE spectra at random
and integrating every second eigenvalue gives a GUE distributed spectrum, and
that integrating out every second eigenvalue of the GSE gives the GOE
\cite{Dy62a, MD63, FR01}. We have not been able to deduce from these relations a LCLT
for the bulk GOE or GSE. But there are other inter-relations between bulk scaled
random matrix
ensembles which are suitable for this purpose \cite{Me92},
\begin{align*}
&E_1(2n;(0,2s)) +  E_1^{\rm bulk}(2n\pm1;(0,2s)) = E^{\pm}(n;(0,s)) \\
&E_4(n;(0,s)) = {1 \over 2} \Big (  E^{+}(n;(0,2s)) + E^{-}(n;(0,2s)) \Big ).
\end{align*}
On the RHSs the superscripts $\pm$ refer to the determinantal point process with
kernels ${1 \over 2} (K_{\rm sin}(x,y) \pm K_{\rm sin}(x,-y) )$, with $K_{\rm sin}$ referring
to the sine kernel, while the subscripts on the LHS refer to the value of $\beta$.
Theorem \ref{T1} applies to $E^\pm(n;(0,2s))$ with $\mu = s$
and $\sigma_\pm^2 \sim {1 \over 2} \sigma^2 |_{\beta = 2}$, so we see immediately
that the bulk scaled GSE satisfies the LCLT (\ref{rK1}) with
$\sigma^2|_{\beta = 4} \sim  \sigma^2|_{\beta = 2} /2$, and that the sum
$E_1(2n;(0,2s)) + E_1(2n\pm 1;(0,2s)) $ satisfies a LCLT.
In particular this latter result implies $ E_1(2n\pm 1;(0,2s)) $ are asymptotically
equal. Combining this  with an anticipated but as yet unproven unimodal property of
$\{E_1(n;(0,s))\}$ would then imply $E_1(2n;(0,2s))$ and 
$E_1(2n\pm 1;(0,2s)) $ are asymptotically equal, and
 the expected LCLT for the individual
$E_1(n;(0,s))$ would follow.

Using Bornemann's package we note that as for the bulk scaled GUE, the finite J
probabilities for the bulk scaled GOE and GSE
are well approximated by a LCLT. This we have done in  Table \ref{T2} for $|J|=10$.
For the corresponding values of $\sigma_J^2$, we have made use of values accurate
up to and including the constant: $\sigma_J^2 \sim (2/(\pi^2 \beta)) \log |J| + B_\beta$
\cite[Eq.~(14.87)]{Fo10}.

\begin{table}
\begin{center}
\begin{tabular}{c||c|c|c|c|c|c|c}
$k$& 7 & 8 & 9 & 10 & 11 & 12 & 13  \\\hline
exact $\beta = 1$ & 0.0027 & 0.0464 & 0.2427 & 0.4169 & 0.2416 & 0.0467 & 0.0029  \\ \hline
Gaussian & 0.0029 & 0.0463 & 0.2413 & 0.4185 & 0.2413 & 0.0463 & 0.0029
\\\hline
exact $\beta = 4$ & $8.65\times 10^{-7}$ & 0.0028 & 0.1819 & 0.6307 & 0.1818 & 0.0028 & $9.7 \times
10^{-7}$ \\ \hline
Gaussian & $5.7 \times
10^{-6}$ & 0.036 & 0.176 & 0.641 & 0.176 & 0.0036 & $5.7 \times
10^{-6}$
\end{tabular}
\caption{\label{T2} Tabulation of $E(k;J)$ for the bulk GOE and GSE, with $|J| = 10$, and the corresponding
Gaussian form. In the latter $\mu_J = 10$ and $\sigma_J = 0.908$ for the GOE, and
$\sigma_J = 0.387$ for the GSE.}
\end{center}
\end{table}


\section{Concluding remarks}
While we have provided a rigorous demonstration of a LCLT for the
bulk scaled GUE, GSE and GOE (the latter for $E_1(2n;(0,2s)) + E_1(2n\pm 1;(0,2s))$),
more generally one expects  LCLT's to hold for  the
$\beta$ generalization of the Gaussian ensembles \cite[\S 1.9]{Fo10}
for general $\beta > 0$. Explicit examples of such LCLT are stated as conjectures in
\cite[Conj.~6]{Fo12a}. For the finite $N$ circular $\beta$-ensemble, and with
$J = (0, \phi)$ a segment of the unit circle, a
CLT  for $N(J)$ in the limit $N \to \infty$ has been established by Killip \cite{Ki08}.
As we have seen, a sufficient condition for a LCLT, which of course implies a CLT,
 is that the zeros of
(\ref{rK5}) are all negative real, together with $\sigma(J_s) \to \infty$ as $s \to \infty$. 
However, as only the case $\beta = 2$ is determinantal, we have no way of establishing such a property, or even providing numerical evidence for general $\beta > 0$. An exception is the
cases $\beta = 1$ and $\beta = 4$ which, as commented in the subsection above relating to the GOE and GSE, have a Pfaffian structure. Making use of this Pfaffian structure,
recent numerical studies have been carried out in \cite{Ka14} which indicate
that the zeros of (\ref{rK5}) for the finite $N$ circular symplectic ensemble are all on the negative
real axis.

Our application of Theorem \ref{T1} has been focussed on random matrices. But
there are other well known examples of determinantal 
processes
 in statistical physics and mathematics obeying the conditions
of the theorem and
which thus must then exhibit the same LCLT (see e.g.~\cite{HKPV08}). A prominent example, conditional upon
the validity of the Montgomery-Odlyzko law, is the Riemann zeros for large modulus
\cite{KS00a}. The Montgomery-Odlyzko law states that certain statistical properties of the latter,
upon appropriate scaling,
coincide with the bulk scaled GUE and this if valid form a determinantal point process.
A proof that  these zeros satisfy a LCLT is an open question, as is the weaker statement of a
CLT (for partial results relating to a smoothed version of the latter, see \cite{HZ02}).

Spin polarized free fermions in dimension $d$ provide examples  of determinantal point processes in higher dimensions \cite{TSZ08}. With $k_F = 2 \sqrt{\pi}(\Gamma(1+d/2))^{1/d}$, the corresponding
bulk scaled (unit density) kernel is computed to equal $c_F J_{d/2}(k_F || \vec{x} - \vec{y} ||)/
(k_F || \vec{x} - \vec{y} ||)^{d/2}$ where $J_\nu(x)$ denotes the usual Bessel function and
$c_F = 2^{d/2} \Gamma(1 + d/2)$. When $d=1$ this corresponds to the sine kernel. From the
explicit form of the kernel substituted in (\ref{JJ}), it is shown in \cite{TSZ08} that for $J$ a sphere
of radius $R$, $\sigma^2_J/R^{d-1}$ is proportional to $\log R$ in the limit $R \to \infty$,
and in particular  $\sigma^2_J$ diverges in this limit so Theorem \ref{T1} applies.

It should also be mentioned that in random matrix theory one encounters 
determinantal point processes in which the correlation kernel is real and non-symmetric.
A simple example is the rank one perturbation of the GUE at the soft edge, 
with parameter tuned so that it
corresponds to the critical regime for the separation of the largest eigenvalue (see e.g.~\cite[eq.~(7.41)]{Fo10}  for the precise form of the kernel). Although Theorem \ref{T1} does not apply directly,
since the eigenvalues of the rank one perturbed matrix strictly interlace those of the
unperturbed one (Cauchy interlacing theorem) we see that
the LCLT is inherited from the corresponding LCLT for the
soft edge GUE. 

\section*{Acknowledgements}
The work of PJF was supported by the Australian Research Council. The work of JLL was
supported by NSF Grant DMR1104500. 
JLL thanks B.~Pittel, D.~Ruelle and particularly E.~Speer for very helpful information
about LCLT.
We thank T.~Spencer and H.-T.~Yau for the invitation to participate in the IAS Princeton
program `Non-equilibrium dynamics and random matrices', thus facilitating the present
collaboration, and we thank H.~Spohn and P.~Sarnak for comments on various drafts.

 \providecommand{\bysame}{\leavevmode\hbox to3em{\hrulefill}\thinspace}
\providecommand{\MR}{\relax\ifhmode\unskip\space\fi MR }
\providecommand{\MRhref}[2]{%
  \href{http://www.ams.org/mathscinet-getitem?mr=#1}{#2}
}
\providecommand{\href}[2]{#2}

\end{document}